# Temporal Cavity Solitons in a Laser-based Microcomb: A Path to a Self-Starting Pulsed Laser without Saturable Absorption

ANTONIO CUTRONA,[1] PIERRE-HENRY HANZARD,[1] MAXWELL ROWLEY,[1] JUAN SEBASTIAN TOTERO-GONGORA,[1] MARCO PECCIANTI,[1] BORIS A. MALOMED,[2,3] GIAN-LUCA OPPO,[4] AND ALESSIA PASQUAZI[1,*]

[1]*Emergent Photonics Lab (Epic), Department of Physics and Astronomy, University of Sussex, Brighton, BN1 9QH, UK*
[2]*Department of Physical Electronics, School of Electrical Engineering, Faculty of Engineering and the Center for Light-Matter Interaction, Tel Aviv University, POB 39040, Ramat Aviv, Tel Aviv, Israel*
[3]*Instituto de Alta Investigación, Universidad de Tarapacá, Casilla 7D, Arica, Chile*
[4]*SUPA, Department of Physics, University of Strathclyde, Glasgow, UK*
*\*a.pasquazi@sussex.ac.uk*

**Abstract:** We theoretically present a design of self-starting operation of microcombs based on laser-cavity solitons in a system composed of a micro-resonator nested in and coupled to an amplifying laser cavity. We demonstrate that it is possible to engineer the modulational-instability gain of the system's zero state to allow the start-up with a well-defined number of robust solitons. The approach can be implemented by using the system parameters, such as the cavity length mismatch and the gain shape, to control the number and repetition rate of the generated solitons. Because the setting does not require saturation of the gain, the results offer an alternative to standard techniques that provide laser mode-locking.

## 1. Introduction

'Microcombs', or optical frequency combs in nonlinear microcavity resonators, are well known as a promising technology for the design of compact metrological sources [1–5]. Leveraging on the rich background of nonlinear dissipative systems in optics [6–10], microcombs find applications also in spectroscopy, optical communications, microwave photonics, and frequency synthesis [11–19].

In particular, microcombs operating with temporal cavity solitons (TCS) have been shown to optimize spectral features of the frequency combs in terms of both the bandwidth and spectral density [20–25]. Such pulses are the temporal counterpart of spatial cavity solitons [10,26–30], i.e., self-confined waves balancing dispersion and nonlinearity in optical resonators while maintaining equilibrium between losses and energy supply. TCS, first predicted in [31] and observed in fiber loops [32], have also been largely studied in microresonators in passive configurations that are well described by the Lugiato-Lefever equation (LLE) [1–3].

In general, these types of localized dissipative waves arise in bistable systems, in parameter regions where an unstable uniform continuous-wave (CW) high-energy state coexists with a stable low-energy one. In such a configuration, cavity solitons can be produced by a local perturbation applied to the low-energy state. Being able to store energy and information, cavity solitons have been extensively studied as a basis for memory schemes, both spatial [26,28] and temporal [33–35] ones. In this context, methods allowing to write and erase cavity solitons locally are well established in both spatial schemes and their low-repetition-rate temporal counterparts. This is usually hard to realize in microcavities, where writing a soliton requires more complex start-up approaches, usually involving a perturbation applied to the background state [2,3].

In this framework, it is relevant to stress that an essential feature of a pulsed optical source is its ability to self-start from noise, being initially turned on from an off state. This has been largely discussed for mode-locked lasers, being configurations such as Kerr-lens mode-locking a primary example of a self-starting laser. In microcombs, this feature is typically demonstrated by delocalized Turing-type "roll solutions". These periodic patterns are generated by the modulational instability (MI) of the low-energy background state, hence they can self-start. Turing rolls usually have a very small number of spectral components in comparison to solitons, the latter species of nonlinear modes being preferable for metrological microcomb applications, while Turing patterns are more suited for microwave technologies [36–38].

Cavity solitons are strictly stable when the underlying low-energy uniform state is stable (solitons over unstable backgrounds can exist, but their existence is limited in conditions and range [39,40]). This property has critical consequences for the occurrence of the start-up of the mode: when the system is initiated in the stable low-energy state, no perturbation can spontaneously grow, thus pulses cannot naturally emerge.

For this reason, much interest in the study of microcombs has been drawn to states such as soliton crystals [41,42], which can start from the MI of the background. The respective solutions may appear as the limit form of the Turing patterns entering the region in the parameter space where solitons are possible [43].

Notably, the MI of the background state can be affected by modifying, for example, the dispersion of the nonlinear gain profile, as in the case of self-injection locking [44]. This has led to the experimental demonstration of self-starting soliton crystals in microcombs [45].

Microcombs with gain elements combined with the nonlinear microcavity [15,21,46–53] have been drawing increasing interest in the course of ongoing studies. In addition to the self-injection locking mechanism for Kerr microresonators [15,21,46–49], cavity solitons have also been observed in an injected quantum-cascade laser [52], for which an LLE-based model has been recently derived [53].

Furthermore, cavity solitons in laser microcombs with a microcavity nested in a gain cavity [52-53] have been recently discovered and explained [54]. Cavity solitons in this setting exist without a background, similarly to laser-cavity solitons in the spatial [29,30,55] and semiconductor-based temporal [34] configurations.

The corresponding model consists of a set of coupled equations, shown in Section 2, which shares important properties to those previously elaborated for the frequency-selected feedback in spatial semiconductor and coupled-fiber amplifiers [56–60]. Two equations are, in general, sufficient to capture the basic features of the system. The model can be expanded by adding additional equations to improve its accuracy in describing the real physical setting.

An important peculiarity of such model, in comparison to previous works for general mode-locked systems, is the absence of fast saturation of the gain: pulses in this system are sustained solely by the Kerr nonlinearity of a nonlinear-Schrödinger-like equation coupled to an equation with linear gain. Usually, gain saturation or, equivalently, nonlinear absorption is required in a laser to generate stable pulses. Laser mode-locking is usually well described by the generalized complex Ginzburg-Landau equation (CGLE), which requires to include a quintic term for producing stable solitons [1,57]. On the other hand, starting from the consideration of the model for fiber-coupled amplifiers [57], it was found that a coupling of the CGLE to a linear equation was sufficient to stabilize the system with the cubic-only nonlinear term, without including any quintic nonlinearity. In this connection, it is relevant to mention that the recent demonstration of laser-cavity soliton microcombs shows that nonlinear gain-attenuating elements are not required in such a coupled system while maintaining stable dissipative solitons [54]. This possibility leads to a considerable reduction in the complexity of the microcomb devices.

Finally, the use of a system of two coupled equations allows one to keep the gain-filtering element separated from the nonlinear Kerr element. Gain-filtering is necessary for any stable amplifying system, allowing it to perform nonlinear broadening of the spectrum without any bandwidth limitation. Thus, this system of coupled equations represents a simple prototype of an alternative type of broadband optical sources completely free of saturation elements.

In this paper, we discuss the ability of the laser TCS system to self-start from noise. We first discuss the stability properties of the solitons in detail, also showing that the system supports a stable analytical solution in the simplest configuration: a nonlinear Schrödinger equation coupled to a dispersion-free equation carrying the linear gain.

Then, we show how to engineer the MI gain of the trivial state to produce a solitary solution that can start directly from noise. A recent study [61] considered the MI spectral gain of the trivial state and its impact on Turing patterns. Here we study how it affects the buildup of solitons and soliton crystals from a noisy zero solution. We find domain boundaries for the trivial solution in a simple analytical form, which allows us to reveal the effects of different laser parameters. The effects of the walk-off and convection have been studied for noise-sustained bidimensional structures, that, however, are short-lived ones [62,63]. In transverse nonlinear optics, gain diffusion has also been shown to modify the stability region of solitary waves [64]. Here we show how to use these parameters to secure the self-start microcomb laser solitons, select the number of solitons in the emerging "crystal" (or array of pulses), control their repetition rate, and extend their existence in the parameter space towards high energies.

## 2. The Single-Mode Nested-Cavity Model with Gain: Linearly Coupled Equations

In its simplest form, the model studied here and related to recently implemented experimental settings [54,61] has the form of a lossy nonlinear Schrödinger equation for complex field $a$, linearly coupled to a lossy linear dispersive equation for the complex field $b$ [56,61,65-67]:

$$\partial_t a = \frac{i\zeta_a}{2}\partial_{xx}a + i\,|a|^2 a \, - \kappa a + \sqrt{\kappa}b, \quad (1)$$

$$\partial_t b = -v\partial_x b + \left(\frac{i\zeta_b}{2} + \sigma\right)\partial_{xx}b + (2\pi\,i\,\Delta + g - 1)b + \sqrt{\kappa}a. \quad (2)$$

The system is written in terms of the propagation variable and longitudinal coordinate (or, in other words, slow and fast times), $t$ and $x$. The equation for $b$ includes the group-velocity and frequency mismatch terms, accounted for by coefficients $v$ and $\Delta$, respectively, with $v$ being often referred to as the walk-off parameter. Coefficients $(g-1)$ and $\sigma$ represent the linear gain/loss and dispersive-loss terms, respectively. In the present scaling, the linear-coupling constant, $\sqrt{\kappa}$, also determines the linear-loss factor, $\kappa$, in the equation for the field $a$. Similar models to Eqs. (1) and (2) have also been discussed in the context of spatial semiconductor frequency-selective feedback lasers [56] and coupled fiber amplifiers [57-60, 68]. Interestingly, Eqs. (1) and (2) are obtained for a configuration with two nested cavities [54,61], which feed energy respectively one another on two different points of the system. For this reason, in this system, the linear coupling terms $\sqrt{\kappa}$ are real and have the same sign, representing a *cross-gain* of the coupled modes. This formulation is different from the coupled system considered in [57-60] and [68], where the modes are conservatively coupled like, for instance, in coupled waveguides. In the latter case, the coupling terms are imaginary.

Group-velocity-dispersion (GVD) coefficients in Eqs. (1) and (2) are $\zeta_{a,b}$. One can consider the case of fully anomalous GVD, with both $\zeta_{a,b}$ being positive, and fully normal GVD, with both $\zeta_{a,b}$ being negative, as well as the mixed case, with opposite signs of $\zeta_a$ and $\zeta_b$. Note that Eq. (1) implies the focusing sign of the Kerr nonlinearity. In general, the cases of the focusing nonlinearity and fully anomalous GVD, and that of a defocusing medium combined with fully normal GVD are mutually equivalent, as the equations for these cases are complex conjugates of each other, apart from the change of sign of the mismatch $\Delta$. This means that there are four main cases of interest. Two of them correspond to the focusing medium and fully anomalous or fully normal GVD, and two others to the same nonlinearity combined with mixed GVDs, that have $\zeta_a > 0, \zeta_b < 0$ or vice versa. Following the experimentally relevant settings [54,61], we concentrate here on the case of the focusing medium and fully anomalous GVD, which makes it possible to construct a bright soliton. Other cases that allow one to produce, e.g., dark solitons, will be presented elsewhere.

## 2.1. Continuous Wave (CW) Solutions

The system of Eqs. (1) and (2) admits the trivial CW solution $a = b = 0$. Taking the linearized version of Eqs. (1) and (2) for $x$-independent solutions, and assuming perturbations in the form of $(\tilde{a}, \tilde{b}) = (\delta a, \delta b) \exp(\lambda t)$, one arrives at the following quadratic equation for the instability gain $\lambda$ of the zero state:

$$\lambda^2 + (1 - g + \kappa - 2\pi i \Delta)\lambda - \kappa(g + 2\pi i \Delta) = 0. \tag{3}$$

A straightforward analysis of this equation demonstrates that it gives rise to instability, i.e. solutions with Re $\lambda > 0$, if $g$ exceeds a minimum threshold value determined by equation

$$(2\pi\Delta)^2 = [(2\pi\Delta)^2 + (1 + \kappa - g_{th})^2] g_{th}. \tag{4}$$

Exactly at $g = g_{th}$, the instability gain takes purely imaginary eigenvalues, $\lambda = \pm i \kappa \sqrt{g_{th}/(1 - g_{th})}$. It follows from Eq. (4) that $g_{th} \approx (2\pi\Delta)^2 (1 + \kappa)^{-2}$ for small mismatch $\Delta$, and $g_{th} \approx 1 - \kappa^2 (2\pi\Delta)^{-2}$ for large values of the mismatch. Under the fact that both these relations predict $0 < g_{th} < 1$, the gain coefficient $g$ in Eq. (2) is assumed to take values between 0 and 1: it must be positive to provide the instability gain, and it must be smaller than 1 (in the present notation) to prevent blowup of field $b(x, t)$.

In addition to the trivial state, the system also admits a set of two CW solutions, which can be obtained by substituting the respective ansatz, $a(t, x) = \sqrt{I} \exp(-i2\pi\phi_A t)$, $b(t, x) = B \exp(-i2\pi\phi_B t)$. It solves the system under the condition $\phi_A = \phi_B = \phi$, which allows finding two solutions [54,61]:

$$\phi_\pm = -\left(\Delta \pm (2\pi)^{-1} \sqrt{(1 - g)g}\right), \tag{5}$$

$$I_\pm = \left(2\pi\Delta \pm \frac{\sqrt{g}(1 - g + \kappa)}{\sqrt{1 - g}}\right). \tag{6}$$

Note that the stationary intensities of the two fields belonging to these solutions are related by $|b|^2 = \kappa (1 - g)^{-1} |a|^2$, while the phases of the fields differ in the two cases, leading to relation $B = \sqrt{\kappa I_\pm} (1 \mp i \sqrt{g(1 - g)^{-1}})$.

There are two threshold values of the mismatch at which field $a$ acquires nonzero intensity:

$$\Delta_{th}^\pm = \mp (2\pi)^{-1} \frac{\sqrt{g}(1 - g + \kappa)}{\sqrt{1 - g}}. \tag{7}$$

As the gain coefficient $g$ belongs to interval $0 < g < 1$, the expression (7) is well defined. For a given value of $g$, when increasing $\Delta$ from negative to positive values, one first passes a first threshold $I_+$, at which the CW solution emerges via a pitchfork bifurcation [69]. Above this threshold, there are two coexisting states, $I_+$ and the zero solution, and the intensity of the branch $I_+$ increases linearly with $\Delta$. Increasing $\Delta$ further, a second threshold is passed, at which the solution $I_-$, with $I_- < I_+$, appears through its bifurcation. Above the second threshold, there are three coexisting solutions: zero, $I_-$ and $I_+$, in order of increasing magnitude.

The thresholds given by Eq. (7) are fully tantamount to the stability boundary of the zero state given by Eq. (3). This conclusion is natural, as nonzero solution branches emerge simultaneously with the onset of the zero state instability.

Knowing the stability conditions for the zero solution helps to identify the (in)stability of the nonzero CW branches. Increasing $\Delta$ across the first bifurcation, as mentioned above, one has two stationary solutions, one of them (the zero state), being unstable above the bifurcation point. This fact implies, in agreement with general principles of the bifurcation theory [69], that the solution branch $I_+$ is stable, as straightforward examinations of Eqs. (1) and (2) demonstrate that the loss terms do not allow the solutions growing to infinite values. Above the second bifurcation point, the zero solution is stable again, hence the new CW solution, $I_-$, with the lower intensity, has to be unstable, while the upper branch $I_+$ remains stable, in the framework of the $x$-independent equations ($I_-$ is unstable as a separatrix between attraction basins of the two stable states). Note, however, that $I_+$ may be affected by MI [61], which can give rise to stable solitons. This is generally expected when the system is bistable, with a modulationally unstable upper branch. This argument points out at a possible existence of a *stability region for solitons* at $\Delta > \Delta_{th}^-$.

## 2.2. Soliton Solutions

Solitons in this system are, in general, not available in an analytical form. Before resorting to numerical analysis, we discuss analytical soliton solutions in a particular region of parameters. A set of analytical Schrödinger-type solutions can be found in the case of a dispersionless, gain-flat amplifier and group-velocity matched system, i.e. $\zeta_b = 0, \nu = 0$ and $\sigma = 0$. In particular, following [57,65,68], it is possible to explore solutions in the form of $a(t,x) = \sqrt{I} \operatorname{sech}(\sqrt{\eta}\, x) \exp(-2\pi i \phi\, t)$, $b(t,x) = B \operatorname{sech}(\sqrt{\eta}\, x) \exp(-2\pi i \phi\, t)$.

Substituting this ansatz in Eqs. (1) and (2), we arrive at a set of simple equations:

$$\left(\left(\frac{i\zeta_a}{2}\eta + 2i\pi\phi - \kappa\right)\sqrt{I} + i(I - \zeta_a\eta)\sqrt{I} \operatorname{sech}^2(\sqrt{\eta}\, x) + \sqrt{\kappa}\, B\right) = 0, \qquad (8)$$

$$\left(\sqrt{\kappa}\sqrt{I} + \left(g - 1 + 2i\pi(\Delta + \phi)\right)B\right) = 0. \qquad (9)$$

A familiar NLS-type linear relation between the intensity and the inverse pulse-width follows from here, $\eta = I/\zeta_a$ (as the coefficient in front of the single $x$-dependent term, $\operatorname{sech}^2(\sqrt{\eta}\, x)$, must vanish), along with formulas that determine two possible solutions:

$$\phi_\pm = -\left(\Delta \pm (2\pi)^{-1}\sqrt{(1-g)g}\right), \qquad (10)$$

$$I_\pm = 2\left(2\pi\Delta \pm \frac{\sqrt{g}(1-g+\kappa)}{\sqrt{1-g}}\right). \qquad (11)$$

Note that the relations for frequency $\phi$ (Eq. 10) and amplitude $B = \sqrt{\kappa I_\pm}\left(1 \mp i\sqrt{g(1-g)^{-1}}\right)$ are the same as for the CW state (see Eq. (5)), while the peak power, given by Eq. (11), is doubled in comparison to its CW counterpart in Eq. (6). Therefore, these soliton solutions share the bifurcation thresholds (see Eq. 7) with the CW states, and similar conjectures can be made for their stability: only the trivial solution exists and is stable at $\Delta < \Delta_{th}^+$, for $\Delta_{th}^+ < \Delta < \Delta_{th}^-$ solution $I_+$ exists along with the unstable trivial state and is unstable too (as a localized state with the unstable zero background), while at $\Delta > \Delta_{th}^-$ there is a solution $I_-$, with lower power than $I_+$, which plays the role of the separatrix between the stable trivial state and a possibly stable upper soliton branch $I_+$ [59,60].

We have verified these conjectures numerically, including numerical solutions for solitons in the case of nonzero coefficients $\nu, \zeta_b$ and $\sigma$ in Eq. (2), in which case the above analytical solutions are not available. To this end, we performed a numerical continuation, starting from the particular analytical solution given by Eqs. (10) and (11), and iteratively advancing into the parameter space from the initial solution using the *pde2path* software package [70]. A set of stationary states produced by the continuation process were identified as branches of the general solution family, and their stability against small perturbations was explored. In particular, we calculated the eigenvalue spectrum of the system's Jacobian at each point of the continuation. This technique made possible to detect instabilities arising from small perturbations with a spatial dependence whose maximum spatial frequency was limited by the discretization of the spatial domain. In Fig. 1, we present the peak intensity of the field $a$, following the variation of mismatch $\Delta$ in Eq. (2) for a set of other parameters ($\kappa = 2\pi$, $\zeta_a = 1.25 \cdot 10^{-5}$, $g = 0.1$), while the GVD coefficients were taken as $\zeta_b = h \times 3.5 \cdot 10^{-5}$ and $\sigma = h \times 1.5 \cdot 10^{-4}$, with $h$ varying from 0 to 1. The objective is to trace a link between the analytical solution and experimentally measured modes reported in [54].

The graphs in Fig. 1 clearly confirm the existence of the stability region at $\Delta > \Delta_{\text{th}}^-$, analytically predicted for the soliton branches in the case of $\nu = \zeta_b = \sigma = 0$. Interestingly, when a nonzero dispersion is added, the branches merge into a single one, which defines the largest possible energy for the soliton, with the top branch maintaining the stability properties in the bistability region, $\Delta > \Delta_{\text{th}}^-$. These results show how the solitons considered in [54] can be produced as an analytical continuation of the nonlinear-Schrödinger-type solitons given by Eqs. (10) and (11). While the field $a$ maintains a well-confined profile under the action of the Kerr nonlinearity, the field $b$ spreads out in the bottom plot of Fig. 1(c), when GVD is included.

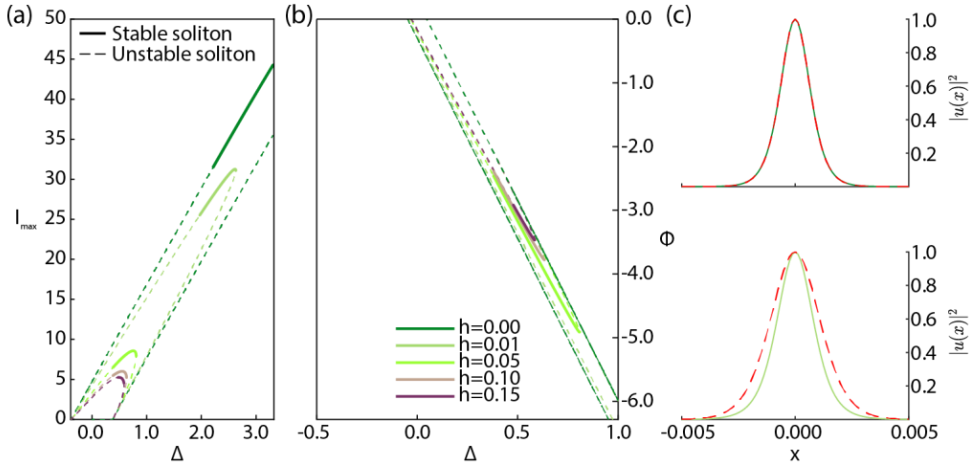

Fig. 1. Stability branches of the soliton states, shown by the dependence of the peak intensity (a) and frequency (b) on mismatch $\Delta$, for a set of other coefficients: $\kappa = 2\pi$, $\zeta_a = 1.25 \cdot 10^{-5}$, $g = 0.1$, and $\zeta_b = h\, 3.5 \cdot 10^{-5}$, $\sigma = h\, 1.5 \cdot 10^{-4}$, with $h$ varying from 0 to 1 (green to purple), $h = 0$ pertaining to the analytical solutions given by Eqs. (10) and (11). Continuous lines represent stable solutions. In panel (c) selected stable solitons are displayed for a specific value of $\Delta$, showing the intensity profiles for fields $a$ and $b$ (both represented by symbol $u$) by solid green and dashed red lines, respectively. The profiles are normalized to the peak values of the intensity. Top and bottom panel in (c) pertain to $h=0$, $\Delta = 2.99$ and $h=0.01$, $\Delta = 2.27$, respectively.

## 2.3. The Modulational-Instability Spectrum of the Zero State and Start-Up from Noise

The study of the MI spectrum of the trivial state allows predicting the parameter ranges corresponding to the existence of stable solitons and exploring the dynamical scenarios through which they arise. The appearance of a spectrum that is unstable against higher spatial frequencies $f_x$ is beneficial for the growth of noise perturbations into multipeaked soliton solutions. Some analytical considerations on the MI spectrum can be done by extending the previous analysis performed for $x$-independent perturbations to $x$-dependent ones $(\tilde{a}, \tilde{b}) = (\delta a, \delta b)\exp(\lambda t + i\omega_x x)$, being $\omega_x = 2\pi f_x$. By deriving an equation for eigenvalue $\lambda$ and setting it to be purely imaginary to identify stability boundaries as done above in Section 2.1 (see Eqs. (3) and (7)), we find two thresholds:

$$\Delta_{\text{th}}^{\pm} = \frac{2\nu\omega_x + (\zeta_b - \zeta_a)\omega_x}{4\pi} \pm \frac{(1 - g + \sigma\omega_x^2 + \kappa)}{2\pi} \sqrt{\frac{(g - \sigma\omega_x^2)}{1 - g + \sigma\omega_x^2}}. \qquad (12)$$

The stability boundary in the plane of the detuning Δ and spatial frequency is better visualized in the implicit form:

$$\left(2\pi\Delta_{\text{th}} - \nu\omega_x - \frac{(\zeta_b - \zeta_a)\omega_x^2}{2}\right)^2 = (1 - G(\omega_x) + \kappa)^2 \frac{G(\omega_x)}{1 - G(\omega_x)}, \qquad (13)$$

where $G(\omega_x)$ is the gain spectral response, which is $g - \sigma\omega_x^2$ in this case, but may be generalized to take into account different gain and filter spectral shapes in more general systems. Note that for $\omega_x = 0$, Eq. (13) reproduces Eq. (7) exactly. For $\nu \to 0$ and small (or properly matched) dispersions $\zeta_{a,b}$, the MI boundary assumes an ellipse-like shape, as determined by the gain-dispersion parameter σ, see Figs. 2(a) and (b). The shape tends to degenerate into straight lines defined by the threshold given by Eq. (7) for $\sigma = 0$. Thus, the gain dispersion allows one to control the maximum MI frequency. Note that, for all values of detuning Δ, the instability region always includes $\omega_x = 0$, which corresponds to the maximum of the MI gain and rules the instability dynamics.

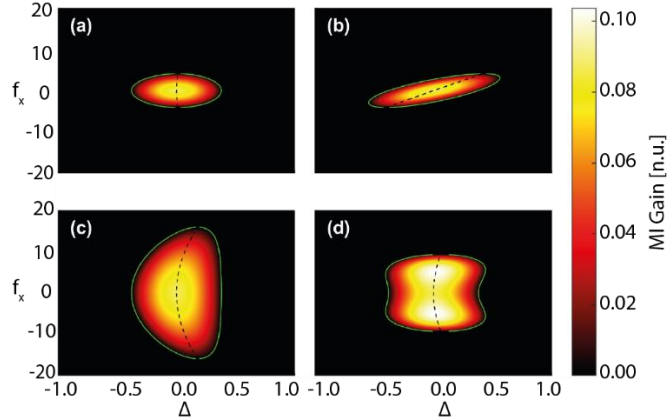

Fig. 2. Maps of the MI gain spectrum of the zero state in the plane of detuning frequency Δ and modulation frequency $f_x$ for Eqs. (1) and (2). Numerically calculated MI gain is represented by the color scale. Instability boundaries predicted by Eq. (12) are shown by green lines, and the symmetry axes of the boundaries, determined by the left-hand side of Eq. (13), is shown by black dashed lines. In all the panels, we fix $\kappa = 2\pi$, $\zeta_a = 1.25 \cdot 10^{-5}$, $\zeta_b = 3.5 \cdot 10^{-5}$ and $g = 0.08$. (a) For the given set of parameters ($\sigma = 1.5 \cdot 10^{-4}, \nu = 0$), the ellipse-type shape is dominant. (b) When the group-velocity mismatch is increased to $\nu = 0.1$, the ellipse is distorted along a straight line. (c) When the gain spectrum is enhanced (σ is decreased to $1 \cdot 10^{-5}$), while maintaining $\nu = 0$, the stability domain expands along the axis of modulation frequency $f_x$, to allow observing the parabolic distortion of the axis under the action of the group-velocity-dispersion mismatch, $(\zeta_b - \zeta_a)$. Finally (d), the case of the fourth-order gain, ($G(\omega_x) = g + \sigma_1\omega_x^2 - \sigma_2\omega_x^4$), with $\sigma_1 = 4 \cdot 10^{-5}, \sigma_2 = 2 \cdot 10^{-8}$, is displayed, showing expansion of the instability domain towards high spatial frequencies at large values of detuning Δ.

Parameters such as the walk-off have been previously used to control spatial multipeaked modes in parametric amplifiers, but in these cases, the solitons are pushed out from the central region of the spatial cavity towards the boundary and they decay [62]. On the other hand, temporal cavity systems are intrinsically periodic. Hence such control only affects the repetition rate of the solitons [54]. Here we see that an increase of the group-velocity, $\nu$, leads to a rotation of the ellipse axis in Fig. 2(b). Interestingly, by acting on the group velocity $\nu$ and gain-dispersion parameter σ, it is possible to engineer the shape of the stability region to obtain a set of values of frequency detuning $\Delta$ at which the instability occurs only at high spatial frequencies. If such a region falls within the region of stability of the solitons, there is a possibility that multipeak self-starting solutions can arise. These regions can also be obtained by adjusting the group-velocity dispersion mismatch $(\zeta_b - \zeta_a)$, which determines the distortion of the symmetry axis of the stability boundaries in the form of a parabola by controlling the sign and magnitude of its curvature in Fig. 2(c). In addition, the spectral shape of the gain, $G(\omega_x)$, may be further adjusted to engineer the boundaries of the instability domain. In Fig. 2(d) we have considered a generalized two-peak gain that can be modeled with a simple quartic polynomial,

$$G(\omega_x) = g + \sigma_1 \omega_x^2 - \sigma_2 \omega_x^4, \tag{14}$$

and will be further explored in Section 3.2. Spatial frequencies that maximize the MI gain determine the perturbation spectral components more likely to grow.

According to the previous considerations, in the case of Fig. 2(a) the CW state is the one most likely to appear ($f_x = 0$ implies no spatial dependence). However, in other cases, the presence of MI gain maxima at $f_x \neq 0$ promotes the growth of perturbations with a spatial period determined by such frequency. Note that, for self-starting Turing patterns, a cascading effect of the MI can be used by seeding the pattern in a modulationally unstable CW, as discussed in [61]), while the self-starting generation of solitons requires the stability of the zero background against perturbations with $f_x = 0$.

Here we present a simple example of soliton start-up, induced by the effect of the walk-off with $\nu = 0.1$ (for its MI spectrum, see Fig. 2(b)). First, it is useful to visualize the MI gain as a function of spatial modulational frequency $f_x$, as shown in Fig. 3(a) for two values of detuning $\Delta$. For every value of $\Delta$, it is possible to extract the frequency values corresponding to the MI gain maxima. The distribution of such values in the $(\Delta, g)$ plane is shown in Fig. 3(b). This plot predicts the spatial frequency that is most likely to appear from the start-up of the system.

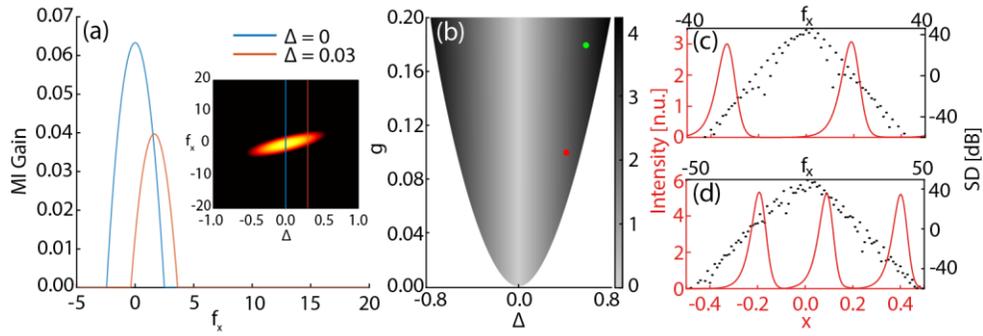

Fig. 3. (a) The MI gain vs. the transverse frequency for $\nu=0.1$, $g=0.08$ with $\Delta=0$ and 0.03 (blue and orange plots, respectively), showing maxima at $f_x=0$ and $f_x=3$. Their position in the MI spectrum plot is shown in the inset by means of the same colors. (b) The distribution of the spatial frequency corresponding to the maximum of the MI gain in the plane of $(\Delta, g)$. Red and green points correspond, respectively, to $\Delta = 0.41$, $g = 0.10$ and $\Delta = 0.58$, $g = 0.18$. (c-d) Soliton regimes (the micro-cavity intensity profile shown in red, and the spectrum shown by black dots) were obtained by calculating the electric field evolution in time from simulations of Eqs. (1) and (2) with a seed of initial noise. The two (three)-soliton regime is reached by choosing the parameters corresponding to the red (green) point in (b). Intensity (red) and spectrum (black) profiles of the micro-cavity electric field are taken from the last time slice.

To secure the start-up of the soliton regime, we also need to know its stability domain, which can be numerically evaluated and, generally, depends on the number of solitons. A particular example of such a calculation is provided in the next section. In general, following the consideration presented in the previous section and Fig. 1(a), solitons are expected to form for positive values of $\Delta$. Here we select two points, namely $\Delta = 0.41, g = 0.10$, and $\Delta = 0.58, g = 0.18$, which are indicated, respectively, in Fig. 3(b) by the red and green points, corresponding to spatial frequencies $f_x = 2.25$ and $f_x = 3.12$. Initiating the propagation with initial white noise and such a set of parameters, we obtain the temporal and spectral outputs displayed in Fig. 3(c) and (d), respectively, where the generation of two and three solitons is evident. The propagation of the three-soliton state is shown in further detail in Fig. 4.

Figures 4(a) and (c) present the temporal evolution in both spatial and spatial-frequency domains. As small white noise is injected in the propagation, the modulation with the spatial frequency $f_x = 3$ starts to grow, as seen in the top plot of Fig. 4(b), eventually transitioning to a triplet of solitons visible in the bottom plot of Fig. 4(b). It is interesting to notice that, after the start-up, the three solitons do not necessarily propagate equally spaced, as can also be seen in Figs. 3(c) and (d). We also remark that all the simulations have been performed over a large number of roundtrips and a broadband frequency noise was added to further confirm their stability.

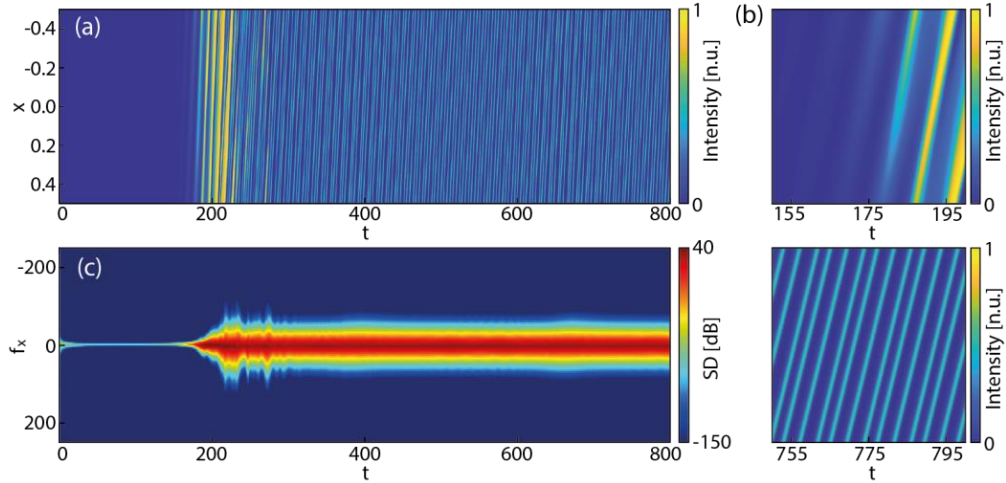

Fig. 4. Self-starting of three solitons with $\Delta = 0.58$, $g = 0.18$, $\nu = 0.1$. Initiated by a noise seed, the time-domain evolution of the model is simulated by means of Eqs. (1) and (2). (a) The evolution of the micro-cavity electric field in the spatial domain. (b) Details of the pattern formation (top) and steady-state propagation (bottom) in the spatial domain. (c) The evolution of the micro-cavity electric field in the spatial-frequency domain.

## 3. The Full Model for the Nested Cavities with Gain

In this section, we focus on the connection of the model with the physical setting and, specifically, with a nonlinear Kerr micro-cavity nested in a main amplifying loop. We introduce a system of mean-field equations derived from the coupled-mode model of the two cavities [54,61]. Here, the field $a$ represents the field in the microcavity, while the field in the main amplifying loop is represented by a superposition of supermodes $b_q$ which are periodic with respect to the microcavity's length. In this way, both $a$ and $b_q$ are functions of the normalized spatial coordinate in the microcavity, $x$, and normalized time, $t$. Further details concerning the physical constants can be found in [54, 61]. The normalized model reads as follows:

$$\partial_t a = \frac{i\zeta_a}{2}\partial_{xx}a + i\,|a|^2 a - \kappa a + \sqrt{\kappa}\sum_{q=-N}^{N} b_q, \quad (15)$$

$$\partial_t b_q = \left(\frac{i\zeta_b}{2} + \sigma\right)\partial_{xx}b_q - \nu\partial_x b_q + 2\pi\,i\,(\Delta - q)b_q + g\,b_q - \sum_{p=-N}^{N} b_p + \sqrt{\kappa}a. \quad (16)$$

Equation (16) governs the evolution of the supermodes identified by index $q$. This model allows to scale up its accuracy by increasing the number $2N+1$ of coupled supermode equations. The case of $N=0$ corresponds to Eqs. (1) and (2) with $b_0 = b$. When the number of amplified modes in the microcavity lines is small (with $\kappa = \pi$ representing one mode per line, and $\kappa = 2\pi$ two modes per line, which is the case addressed by the numerical results reported below), Eqs. (1) and (2) grasp the main physical purport of the solitons. However, it is crucial to take into account the periodic nature of the system. Periodicity directly affects the instability region of the trivial state reported in Fig. 5 for the same parameters as in Fig. 2, where the periodic repetition of the instability areas for multiple values of detuning $\Delta$ is clearly visible. For better accuracy, here and below, we focus on the case $N = 7$, which uses 15 supermodes, allowing us to maintain good fidelity in the description of the system.

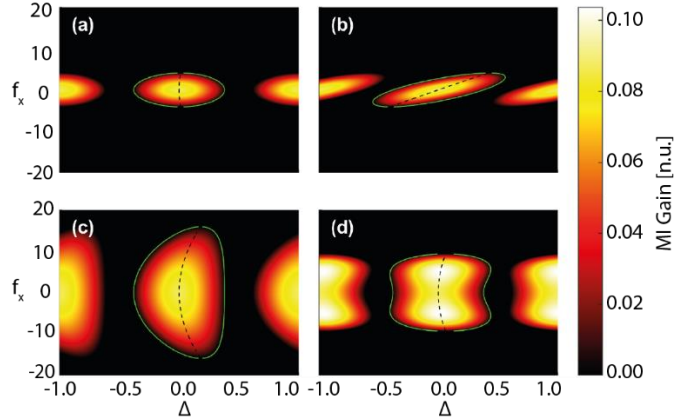

Fig. 5. MI gain spectrum of the zero state in the plane of detuning $\Delta$ and spatial modulation frequency $f_x$, as obtained from Eqs. (15) and (16) with $N = 7$. Numerically calculated values of the gain are displayed by means of the color scale. Instability boundaries analytically predicted by Eq. (12) are shown by green lines. The symmetry axis of the boundary determined by the left-hand side of Eq. (13) is shown by black dashed lines. Parameters are the same as in Fig. (2).

In light of such multiplicity, in the following subsection, we extend the considerations presented above in Section 2 through a detailed numerical study of the variation of the stability region with a change of the walk-off parameter, demonstrating the ability to control the multiplicity of the solitons self-starting from noise.

*3.1. Soliton Stationary States and Start-up from Noise: The Role of Walk-off*

We start by exploring the stability of single-soliton states using a numerical-continuation [70] of Eqs. (15) and (16). Figure 6(a) reports a set of branches obtained by varying the frequency detuning $\Delta$ for a set of gain values $g$. Here the group velocity mismatch equals zero, $\nu = 0$. It is useful to project the stability region of the soliton onto the stability domain of the zero state in the plane of $(\Delta, g)$, where the spatial frequency corresponding to the maximum MI gain is $f_x = 0$. The projection is displayed in Fig. 6(b). The periodicity of the system imposes an upper boundary on the MI gain. Such a stability region can be computed by varying the walk-off parameter $\nu$. The results of the systematic study are reported in Fig. 6(c), where we project the stability regions of the branches obtained at different values of $\nu$ onto the plane of $(\Delta, g)$. The three-dimensional region enclosed by the obtained surface comprises all parameter values supporting the stable single-soliton operation. Here, it is evident that the single-soliton stability condition constrains $\nu$ to be limited to a region of width $\approx 0.24$, considering the symmetry of the stability boundaries about the $\nu = 0$ plane. The stability region for the case $\nu = 0.1$, along with the spatial frequency distribution corresponding to the maximum MI gain is presented in Fig. 6(d). It can be observed that, although an increase of $\nu$ leads to contraction of the stability region in the $(\Delta, g)$ plane, it also helps it to penetrate the instability region of the zero state. In this particular case, it overlaps with the zone where the zero-state's MI gain has a maximum for $f_x = 1$. The zero state is unstable here against perturbations with the spatial period covering the entire transverse range, hence very broad solitons can be created by such a perturbation.

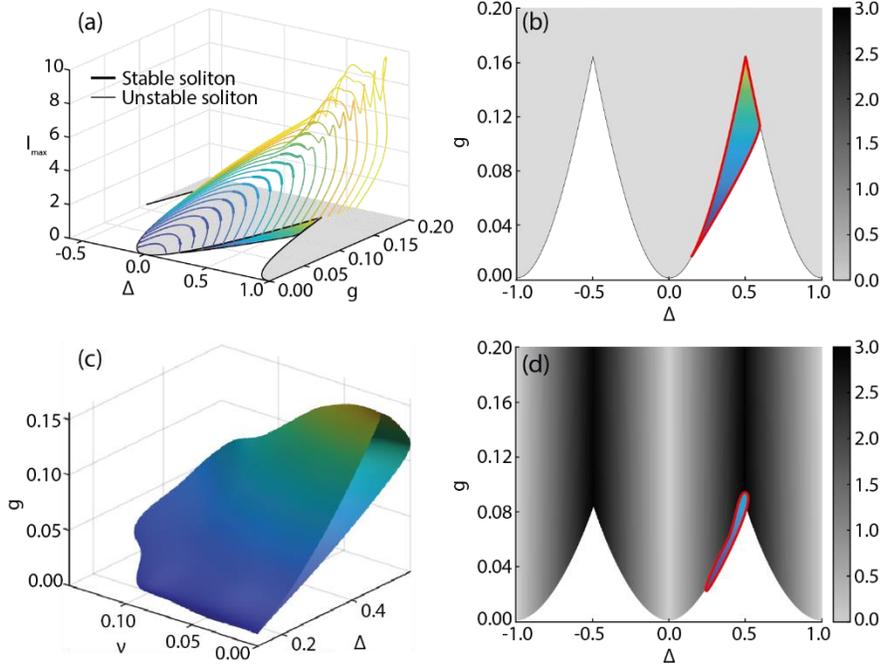

Fig. 6. (a) One-soliton branches shown in terms of the micro-cavity peak intensity at $\nu=0$. Thin and bold lines denote for stable and unstable solitons, respectively, while colors from blue to yellow represent the gain scale. The stability region (blue to yellow colored area) of the solitons is projected on the plane of $(\Delta, g)$. The gray-color plot shows the distribution of the frequency of the maximum MI gain of the zero state. (b) The view of the projected stability region of laser TCS in the plane of $(\Delta, g)$, reported in (a) for $\nu = 0$. (c) The summary of the stability region of laser TCS in the space of $(\Delta, \nu, g)$. The symmetric counterpart of this region is not shown for $\nu < 0$. (d) The same as in (b), but for $\nu = 0.1$.

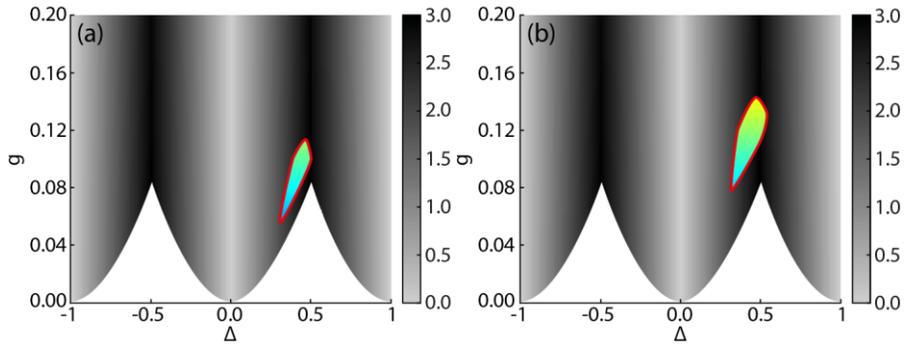

Fig. 7. Soliton stability region for $\nu = 0.1$, along with gray-coded distribution of the spatial frequency corresponding to the maximum MI gain of the zero state, in the plane of $(\Delta, g)$, for two- and three-solitons modes are displayed in panels (a) and (b), respectively. These regions are centered around the instability regions of the zero state with the maximum MI gain attained at $f_x = 2$ and $f_x = 3$ respectively.

Because the MI gain has a spectrum spread in the spatial frequency, we can, furthermore, predict that it affects the stability of multipeak soliton arrays [68] or an equally spaced soliton crystal. We have then reproduced the study of the stationary states for two and three equidistant solitons. Such solutions do not necessarily inherit the stability properties of the underlying one-soliton state, as is evident in Figs. 7(a) and (b). In fact, on the contrary to the single-soliton case, the projection of the stability region onto the $(\Delta, g)$ plane does not overlap with the stability boundaries of the zero state. In particular, two and three equidistantly placed soliton solutions are stable when the spatial frequency corresponding to the largest MI gain takes values around $2 - 3$. This property of the multi-soliton states helps to expand the stability boundaries towards larger values of gain $g$, and thus create high-energy states and broad patterns that are relevant for practical applications.

These considerations provide clear guidance for the engineering of the start-up region of the system. As an example, we realized the start-up for parameters in the stability regions defined in Figs. 6(d) and 7 by injecting low-energy, random noise in the first time slice of the simulations. Figure 8 summarises the self-starting of the one-, two-, and three-soliton states. In each case, we performed the same simulation for different realizations of the initial noise to estimate our strategy's accuracy to control the multiplicity of the produced state. Solutions with a large number of solitons generally correspond to broader spectra, as they appear under the action of a stronger gain, as evident in Fig. 7. Therefore, this approach provides access to stable structures at large values of the gain, which produce only instability in the single-soliton sector, for $\nu = 0$.

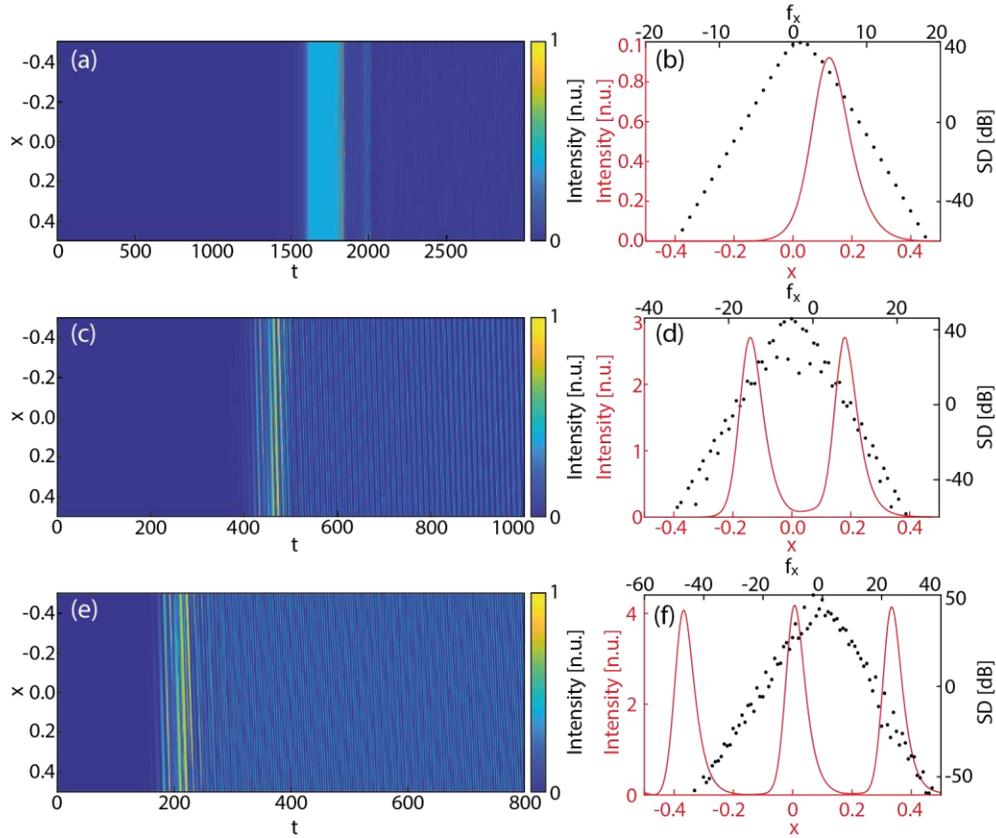

Fig. 8. The self-start of multiple solitons. Initiated by initial noise, the time-domain evolution of the model is produced by simulations of Eqs. (15) and (16), with $N = 7$. (a) The evolution of the micro-cavity electric field in the spatial domain; (b) intensity (red) and spectral (black) profiles

of the micro-cavity electric field in the last time slice for the single-soliton solution ($\Delta= 0.358$, $g = 0.05$, $\nu = 0.1$). Panels (c) and (d) display the same for the two-soliton solution ($\Delta= 0.420$, $g = 0.09$, $\nu = 0.1$). Panels (e) and (f) display the same for three-soliton solution ($\Delta= 0.45$, $g = 0.09$, $\nu = 0.14$).

### 3.2 Engineering the Spectral Gain for Self-starting Multi-soliton Solutions

The MI analysis of the zero state suggests that the particular frequency profile of the gain included in the model plays an essential role in the selection of the spatial frequency corresponding to the peak of the MI gain. In previous sections, we considered a gain spectrum defined as $G(\omega_x) = g - \sigma\omega_x^2$, which is parabolic and centered at $\omega_x = 0$, as is common to most experimental settings. However, as the spectral shape of the gain can be easily engineered through appropriately designed spectral filters, it is relevant to explore the degrees of freedom offered by different spectral profiles. In particular, it is interesting to reproduce the spectrum of a two-peaked gain, which is easily achievable in actual experiments. For simplicity, we consider the symmetric shape modeled by Eq. (14), where the first and third-order terms are absent as we consider the gain centered at $\omega_x = 0$.

We here address the MI induced by such gain spectra. To this end, we extended the system of Eqs. (15) and (16) by adding in the latter equation an extra fourth-order derivative in $x$ with new parameters $\sigma_1$ and $\sigma_2$ obtaining:

$$\partial_t b_q = -\sigma_2 \partial_{xxxx} b_q + \left(\frac{i\zeta_b}{2} - \sigma_1\right)\partial_{xx}b_q - \nu\partial_x b_q + 2\pi i(\Delta - q)b_q$$

$$+ g\, b_q - \sum_{p=-N}^{N} b_p + \sqrt{\kappa}a. \tag{17}$$

The parameter $g$ still refers to the gain at the spatial frequency $\omega_x = 0$ that we keep with a constant difference to the side-lobe peak gain. In Fig. 9(a), the MI gain attains a maximum at spatial frequencies close to the side-lobe peaks for most values of $\Delta$. This is confirmed in Fig. 9(c) by the peak-frequency chart in the $(\Delta, g)$ parameter plane. With the increase of $\nu$, the MI gain frequency profile varies similarly to Fig. 9(b). The variation leads to overlap of adjacent frequency profiles, reducing the stability range of the zero background to a very limited range in the $(\Delta, g)$ plane in Figs. 9(b) and (d). Note that the peak-frequency chart exhibits a sharp change around $\Delta = 0$, with the frequency corresponding to the largest MI gain being $|f_x| = 5$ around $\Delta = 0$.

This fact suggests that the gain-shaping approach effectively extends the range in which the system can self-start with the desired soliton multiplicity, thus offering better control of the soliton number in the emerging patterns in an extensive range of values of frequency detuning $\Delta$. Similar to our previous analysis, these considerations have been validated against time-domain simulations with the noise input (see Fig. 10), confirming that a properly chosen gain-shaping profile forces the system to self-start with the intended number of solitons (five, in that case).

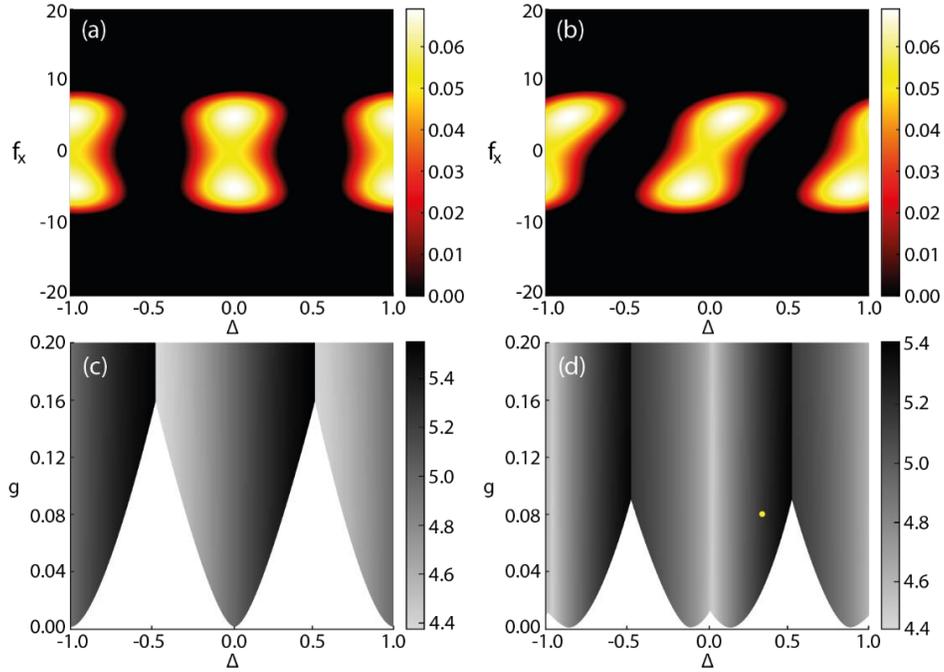

Fig. 9. Effect of the gain shaping, in the form of $G(\omega_x) = g + \sigma_1 \omega_x^2 - \sigma_2 \omega_x^4$, with $\sigma_1 = 4 \cdot 10^{-5}, \sigma_2 = 2 \cdot 10^{-8}$, on the MI gain of the zero state. (a) The MI gain vs. $\Delta$ and the spatial frequency of the modulational perturbation at $\nu = 0$, $g = 0.08$. (b) The same at $\nu = 0.025$, $g = 0.08$. (c) The map of the spatial frequency corresponding to the largest MI gain in the plane of $(g, \Delta)$ at $\nu = 0$. (d) The same at $\nu = 0.025$. Yellow point corresponds to $\Delta = 0.34$, $g = 0.08$.

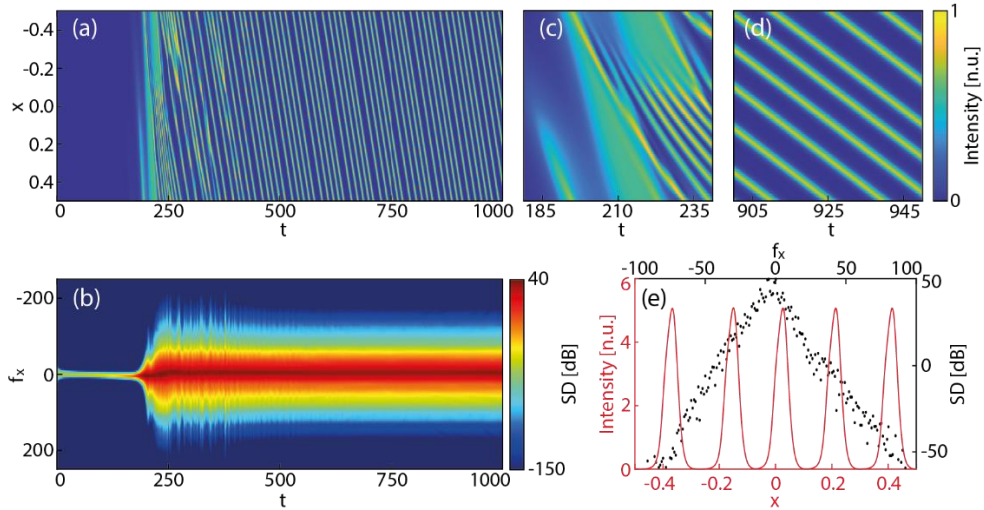

Fig. 10. The multi-soliton self-start, implemented with the shaped gain ($\Delta = 0.34$, $g = 0.08$, $\nu = 0.025$). Initiated by the initial-noise seed, the time-domain evolution of the gain-shaped model is simulated, using Eqs. (15) and (17) with $N = 7$. (a) The evolution of the micro-cavity electric field in the spatial domain. (b) The evolution of the micro-cavity electric field in the spatial-frequency domain. (c) The corresponding pattern formation. (d) The steady-state propagation. (e) Intensity (red) and spectrum (black) profiles in the last time slice.

## 4. Conclusions

We have analyzed properties of laser TCS in the model of a gain-cavity with a nested microcavity where cross-gain terms linearly couple the fields. An essential feature of this model is the absence of nonlinear saturation of the gain, in stark difference with standard models based on CGLEs (Complex Ginzburg-Landau Equations). The analysis focuses on the stability of soliton solutions and on strategies to obtain self-starting regimes, which are crucially essential properties for developing optical sources.

We started with analytical considerations. They result in two distinct analytical solutions for solitary pulses in the model with the dispersion-free amplifier, the higher-amplitude branch being stable for values of the mismatch (detuning frequency), $\Delta$ in Eq. (2), which exceeds the threshold value. It corresponds to the red-detuned frequency in the Kerr-focusing component (Eq. (1)), in agreement with experimental observations [54]. By numerically continuing the original analytical solutions into the experimentally relevant region of parameters, including the gain dispersion, it is possible to produce soliton solutions fitting the experiments. The limitation on the gain bandwidth imposes an upper limit on the maximum energy that the solitons may have for a given gain, but it does not bound the bandwidth of the solitons, nor it affects the stability region. This region is related to the MI (modulational instability) domain of the zero state. We have presented a detailed analysis of the MI gain, deriving analytical forms for the instability boundaries in Eqs. (4), (7), (12) and (13). Then, we addressed different strategies to engineer such a boundary, aiming to identify a set of parameters for which the zero state is unstable to perturbations of large spatial frequencies but stable against spatially-uniform perturbations, for detuning frequencies $\Delta > 0$. It is more likely to obtain stable solitons under such conditions. By exploiting system parameters such as the cavity length mismatch and the gain shape, we observe self-start from noise of single or clusters of laser TCS in the system of Eqs. (1) and (2). Quite notably, our strategy grants direct control over the number and repetition rate of the generated laser TCS.

We have extended the numerical analysis to the full set of Eqs. (15) and (16), which describes more accurately the multimodal structure of the experimental setting. This system also makes it possible to identify self-start regimes from noise and produce chains composed of a required number of solitons by adjusting the group-velocity mismatch, alias the walk-off and the gain-shaping profile, which are readily controllable parameters in an actual experiment.

Although parameters used in this work are focused on modeling a particular experimental setting, the results offer the basis for the development of ultrafast optical-soliton sources with amplifiers that do not use any nonlinear dependence of the gain/loss elements. The stabilization of the pulses is provided by the linear coupling of the two underlying equations, and the system can self-start from noise, producing stable trains of solitons in a large parameter region by taking advantage of the walk-off. The results can be easily understood with the help of the straightforward MI considerations. This work paves the way for designing a new class of optical sources based on coupled waveguides or multiple cavities. As stressed above, they will be stabilized by the linear coupling instead of the usually assumed ultrafast saturation, offering a well-defined operational region to create a controllable number of robust solitary pulses.

**Acknowledgment**

We acknowledge the support of the EPSRC, Industrial Innovation Fellowship Programme, under Grant EP/S001018/1. This project has received funding from the European Research Council (ERC) under the European Union's Horizon 2020 research and innovation programme Grant agreement n° 851758 (TELSCOMBE). AC acknowledges the support of the DSTL-Defence Science & Technology Laboratory through the studentship DSTLX1000142078. JSTG acknowledges the Leverhulme Trust (Leverhulme Early Career Fellowship ECF-2020-537). The work of BAM is supported, in part, by the Israel Science Foundation, through grant No. 1286/27.

## Disclosures

The authors declare no conflicts of interest.

## References

1. L. A. Lugiato, F. Prati, and M. Brambilla, *Nonlinear Optical Systems* (Cambridge University Press, 2015).
2. T. J. Kippenberg, A. L. Gaeta, M. Lipson, and M. L. Gorodetsky, "Dissipative Kerr solitons in optical microresonators," Science **361**(6402), eaan8083 (2018).
3. A. Pasquazi, M. Peccianti, L. Razzari, D. J. Moss, S. Coen, M. Erkintalo, Y. K. Chembo, T. Hansson, S. Wabnitz, P. Del'Haye, X. Xue, A. M. Weiner, and R. Morandotti, "Micro-combs: A novel generation of optical sources," Phys. Rep. **729**, 1–81 (2018).
4. A. A. Savchenkov, A. B. Matsko, and L. Maleki, "On Frequency Combs in Monolithic Resonators," Nanophotonics **5**(2), 363–391 (2016).
5. S. Barland, S. Coen, M. Erkintalo, M. Giudici, J. Javaloyes, and S. Murdoch, "Temporal localized structures in optical resonators," Adv. Phys. X **2**(3), 496–517 (2017).
6. D. W. M. Laughlin, J. V. Moloney, and A. C. Newell, "Solitary Waves as Fixed Points of Infinite-Dimensional Maps in an Optical Bistable Ring Cavity," Phys. Rev. Lett. **51**(2), 75–78 (1983).
7. T. Ackemann, William J. Firth, and G. Oppo, "Chapter 6, Fundamentals and Applications of Spatial Dissipative Solitons in Photonic Devices," in *Advances In Atomic, Molecular, and Optical Physics* (Elsevier, 2009), **57**, 323–421.
8. F. T. Arecchi, S. Boccaletti, and P. Ramazza, "Pattern formation and competition in nonlinear optics," Phys. Rep. **318**(1-2), 1-83 (1999).
9. N. N. Rosanov, *Spatial Hysteresis and Optical Patterns* (Springer, 2002).
10. W. J. Firth and G. K. Harkness, "Cavity Solitons," Asian J. Phys. **7**, 665–677 (1998).
11. P. Marin-Palomo, J. N. Kemal, M. Karpov, A. Kordts, J. Pfeifle, M. H. P. Pfeiffer, P. Trocha, S. Wolf, V. Brasch, M. H. Anderson, R. Rosenberger, K. Vijayan, W. Freude, T. J. Kippenberg, and C. Koos, "Microresonator-based solitons for massively parallel coherent optical communications," Nature **546**(7657), 274–279 (2017).
12. M.-G. Suh and K. J. Vahala, "Soliton microcomb range measurement," Science **359**(6378), 884–887 (2018).
13. V. Brasch, M. Geiselmann, T. Herr, G. Lihachev, M. H. Pfeiffer, M. L. Gorodetsky, and T. J. Kippenberg, "Photonic chip–based optical frequency comb using soliton Cherenkov radiation," Science **351**(6271), 357–360 (2016).
14. D. T. Spencer, T. Drake, T. C. Briles, J. Stone, L. C. Sinclair, C. Fredrick, Q. Li, D. Westly, B. R. Ilic, A. Bluestone, N. Volet, T. Komljenovic, L. Chang, S. H. Lee, D. Y. Oh, M.-G. Suh, K. Y. Yang, M. H. P. Pfeiffer, T. J. Kippenberg, E. Norberg, L. Theogarajan, K. Vahala, N. R. Newbury, K. Srinivasan, J. E. Bowers, S. A. Diddams, and S. B. Papp, "An optical-frequency synthesizer using integrated photonics," Nature **557**(7703), 81–88 (2018).
15. W. Liang, D. Eliyahu, V. S. Ilchenko, A. A. Savchenkov, A. B. Matsko, D. Seidel, and L. Maleki, "High spectral purity Kerr frequency comb radio frequency photonic oscillator," Nat. Commun. **6**(1), 7957 (2015).
16. S.-W. Huang, J. Yang, M. Yu, B. H. McGuyer, D.-L. Kwong, T. Zelevinsky, and C. W. Wong, "A broadband chip-scale optical frequency synthesizer at 2.7 x 10-16 relative uncertainty," Sci. Adv. **2**(4), e1501489 (2016).
17. M. Yu, Y. Okawachi, A. G. Griffith, N. Picqué, M. Lipson, and A. L. Gaeta, "Silicon-chip-based mid-infrared dual-comb spectroscopy," Nat. Commun. **9**(1), 1869 (2018).
18. A. Dutt, C. Joshi, X. Ji, J. Cardenas, Y. Okawachi, K. Luke, A. L. Gaeta, and M. Lipson, "On-chip dual-comb source for spectroscopy," Sci. Adv. **4**(3), e1701858 (2018).
19. J. Pfeifle, A. Coillet, R. Henriet, K. Saleh, P. Schindler, C. Weimann, W. Freude, I. V. Balakireva, L. Larger, C. Koos, and Y. K. Chembo, "Optimally Coherent Kerr Combs Generated with Crystalline Whispering Gallery Mode Resonators for Ultrahigh Capacity Fiber Communications," Phys. Rev. Lett. **114**(9), 093902 (2015).
20. T. Herr, V. Brasch, J. D. Jost, C. Y. Wang, N. M. Kondratiev, M. L. Gorodetsky, and T. J. Kippenberg, "Temporal solitons in optical microresonators," Nat. Photonics **8**(2), 145–152 (2013).
21. B. Stern, X. Ji, Y. Okawachi, A. L. Gaeta, and M. Lipson, "Battery-operated integrated frequency comb generator," Nature **562**(7727), 401–408 (2018).
22. E. Obrzud, S. Lecomte, and T. Herr, "Temporal solitons in microresonators driven by optical pulses," Nat. Photonics **11**(9), 600–607 (2017).
23. W. Liang, A. A. Savchenkov, Z. Xie, J. F. McMillan, J. Burkhart, V. S. Ilchenko, C. W. Wong, A. B. Matsko, and L. Maleki, "Miniature multioctave light source based on a monolithic microcavity," Optica **2**(1), 40–47 (2015).
24. L. Maleki, V. S. Ilchenko, and A. B. Matsko, *Practical Applications of Microresonators in Optics and Photonics* (CRC Press, 2018).
25. S. Zhang, J. M. Silver, L. Del Bino, F. Copie, M. T. M. Woodley, G. N. Ghalanos, A. Ø. Svela, N. Moroney, and P. Del'Haye, "Sub-milliwatt-level microresonator solitons with extended access range using an auxiliary laser," Optica **6**(2), 206–212 (2019).
26. S. Barland, J. R. Tredicce, M. Brambilla, L. A. Lugiato, S. Balle, M. Giudici, T. Maggipinto, L. Spinelli, G. Tissoni, T. Knodl, M. Miller, and R. Jäger, "Cavity solitons as pixels in semiconductor microcavities," Nature **419**(6908), 699–702 (2002).

27. Radwell, N., Mcintyre, C., Scroggie, A. J., Oppo, G-L., Firth, W. J., & Ackemann, T. (2010). Switching spatial dissipative solitons in a VCSEL with frequency selective feedback. European Physical Journal D: Atomic, Molecular, Optical and Plasma Physics, 59(1), 121-131. https://doi.org/10.1140/epjd/e2010-00124-6
28. G. S. McDonald and W. J. Firth, "Spatial solitary-wave optical memory," J. Opt. Soc. Am. B **7**(7), 1328–1335 (1990).
29. Y. Tanguy, T. Ackemann, W. J. Firth, and R. Jäger, "Realization of a Semiconductor-Based Cavity Soliton Laser," Phys. Rev. Lett. **100**(1), 013907 (2008).
30. P. Genevet, S. Barland, M. Giudici, and J. R. Tredicce, "Cavity soliton laser based on mutually coupled semiconductor microresonators", Phys. Rev. Lett. **101**(12), 123905 (2008).
31. M. Haelterman, S. Trillo, and S. Wabnitz, "Dissipative modulation instability in a nonlinear dispersive ring cavity," Opt. Commun. **91**(5–6), 401–407 (1992).
32. F. Leo, S. Coen, P. Kockaert, S.-P. Gorza, P. Emplit, and M. Haelterman, "Temporal cavity solitons in one-dimensional Kerr media as bits in an all-optical buffer," Nat. Photonics **4**(7), 471–476 (2010).
33. M. Marconi, J. Javaloyes, S. Balle, and M. Giudici, "How Lasing Localized Structures Evolve out of Passive Mode Locking," Phys. Rev. Lett. **112**(22), 223901 (2014).
34. M. Marconi, J. Javaloyes, S. Barland, S. Balle, and M. Giudici, "Vectorial dissipative solitons in vertical-cavity surface-emitting lasers with delays," Nat. Photonics **9**(7), 450–455 (2015).
35. J. K. Jang, M. Erkintalo, S. G. Murdoch, and S. Coen, "Observation of dispersive wave emission by temporal cavity solitons," Opt. Lett. **39**(19), 5503–5506 (2014).
36. A. Coillet and Y. Chembo, "On the robustness of phase locking in Kerr optical frequency combs," Opt. Lett. **39**(6), 1529-1532 (2014).
37. S.-W. Huang, J. Yang, S.-H. Yang, M. Yu, D.-L. Kwong, T. Zelevinsky, M. Jarrahi, and C. W. Wong, "Globally Stable Microresonator Turing Pattern Formation for Coherent High-Power THz Radiation On-Chip," Phys. Rev. X **7**(4), 041002 (2017).
38. S. Diallo and Y. K. Chembo, "Optimization of primary Kerr optical frequency combs for tunable microwave generation," Opt. Lett. **42**(18), 3522–3525 (2017).
39. M. Pesch, E. Große Westhoff, T. Ackemann, and W. Lange, "Observation of a Discrete Family of Dissipative Solitons in a Nonlinear Optical System," Phys. Rev. Lett. **95**(14), 143906 (2005).
40. M. Pesch, W. Lange, D. Gomila, T. Ackemann, W. J. Firth, and G.-L. Oppo, "Two-Dimensional Front Dynamics and Spatial Solitons in a Nonlinear Optical System," Phys. Rev. Lett. **99**(15), 153902 (2007).
41. B. Corcoran, M. Tan, X. Xu, A. Boes, J. Wu, T. G. Nguyen, S. T. Chu, B. E. Little, R. Morandotti, A. Mitchell, and D. J. Moss, "Ultra-dense optical data transmission over standard fibre with a single chip source," Nat. Commun. **11**(1), 2568 (2020).
42. D. C. Cole, E. S. Lamb, P. Del'Haye, S. A. Diddams, and S. B. Papp, "Soliton crystals in Kerr resonators," Nat. Photonics **11**(10), 671–676 (2017).
43. Z. Qi, S. Wang, J. Jaramillo-Villegas, M. Qi, A. M. Weiner, G. D'Aguanno, T. F. Carruthers, and C. R. Menyuk, "Dissipative cnoidal waves (Turing rolls) and the soliton limit in microring resonators," Optica **6**(9), 1220–1232 (2019).
44. N. M. Kondratiev and V. E. Lobanov, "Modulational instability and frequency combs in whispering-gallery-mode microresonators with backscattering," Phys. Rev. A **101**(1), 013816 (2020).
45. B. Shen, L. Chang, J. Liu, H. Wang, Q.-F. Yang, C. Xiang, R. N. Wang, J. He, T. Liu, W. Xie, J. Guo, D. Kinghorn, L. Wu, Q.-X. Ji, T. J. Kippenberg, K. Vahala, and J. E. Bowers, "Integrated turnkey soliton microcombs," Nature **582**(7812), 365–369 (2020).
46. W. Liang, V. S. Ilchenko, A. A. Savchenkov, A. B. Matsko, D. Seidel, and L. Maleki, "Whispering-gallery-mode-resonator-based ultranarrow linewidth external-cavity semiconductor laser," Opt. Lett. **35**(16), 2822-2824 (2010).
47. N. G. Pavlov, S. Koptyaev, G. V. Lihachev, A. S. Voloshin, A. S. Gorodnitskiy, M. V. Ryabko, S. V. Polonsky, and M. L. Gorodetsky, "Narrow-linewidth lasing and soliton Kerr microcombs with ordinary laser diodes," Nat. Photonics **12**(11), 694–698 (2018).
48. P. Del'Haye, K. Beha, S. B. Papp, and S. A. Diddams, "Self-Injection Locking and Phase-Locked States in Microresonator-Based Optical Frequency Combs," Phys. Rev. Lett. **112**(4), 043905 (2014).
49. A. S. Voloshin, J. Liu, N. M. Kondratiev, G. V. Lihachev, T. J. Kippenberg, and I. A. Bilenko, "Dynamics of soliton self-injection locking in a photonic chip-based microresonator," arXiv:1912.11303 [physics.optics], (2020).
50. M. Peccianti, A. Pasquazi, Y. Park, B. E. Little, S. T. Chu, D. J. Moss, and R. Morandotti, "Demonstration of a stable ultrafast laser based on a nonlinear microcavity," Nat. Commun. **3**(1), 765 (2012).
51. M. Rowley, B. Wetzel, L. D. Lauro, T. Gongora, J. Silver, L. Del, P. D. Haye, M. Peccianti, and A. Pasquazi, "Thermo-optical pulsing in a microresonator filtered fiber-laser: a route towards all-optical control and synchronization," Opt. Express **27**(14), 19242-19254 (2019).
52. M. Piccardo, B. Schwarz, D. Kazakov, M. Beiser, N. Opačak, Y. Wang, S. Jha, J. Hillbrand, M. Tamagnone, W. T. Chen, A. Y. Zhu, L. L. Columbo, A. Belyanin, and F. Capasso, "Frequency combs induced by phase turbulence," Nature **582**(7812), 360–364 (2020).
53. L. Columbo, M. Piccardo, F. Prati, L. A. Lugiato, , M. Brambilla, A. Gatti, C. Silvestri, M. Gioannini, N. Opacak, B. Schwarz, and F. Capasso, "Unifying frequency combs in active and passive cavities: CW driving of temporal solitons in ring lasers," ArXiv:2007.07533 [physics.optics] (2020).

54. H. Bao, A. Cooper, M. Rowley, L. Di Lauro, J. S. Totero Gongora, S. T. Chu, B. E. Little, G.-L. Oppo, R. Morandotti, D. J. Moss, B. Wetzel, M. Peccianti, and A. Pasquazi, "Laser cavity-soliton microcombs," Nat. Photonics **13**(6), 384-389 (2019).
55. P. Genevet, S. Barland, M. Giudici, and J. R. Tredicce, "Bistable and addressable localized vortices in semiconductor lasers," Phys. Rev. Lett. **104**(22), 223902 (2010).
56. A. J. Scroggie, W. J. Firth, and G.-L. Oppo, "Cavity-soliton laser with frequency-selective feedback," Phys. Rev. A **80**(1), 013829 (2009).
57. B. A. Malomed, "Solitary pulses in linearly coupled Ginzburg-Landau equations," Chaos Interdiscip. J. Nonlinear Sci. **17**(3), 037117 (2007).
58. A. Sigler, B. A. Malomed, and D. V. Skryabin, "Localized states in a triangular set of linearly coupled complex Ginzburg-Landau equations," Phys. Rev. E **74**(6), 066604 (2006).
59. B. A. Malomed and H. G. Winful, "Stable solitons in two-component active systems," Phys. Rev. E **53**(5), 5365 (1996).
60. J. Atai and B. A. Malomed, "Stability and interactions of solitons in two-component active systems," Phys. Rev. E **54**(4), 4371–4374 (1996).
61. H. Bao, L. Olivieri, M. Rowley, S. T. Chu, B. E. Little, R. Morandotti, D. J. Moss, J. S. Totero Gongora, M. Peccianti, and A. Pasquazi, "Turing patterns in a fiber laser with a nested microresonator: Robust and controllable microcomb generation," Phys. Rev. Res. **2**(2), 023395 (2020).
62. M. Santagiustina, P. Colet, M. San Miguel, and D. Walgraef, "Two-dimensional noise-sustained structures in optical parametric oscillators," Phys. Rev. E **58**(3), 3843–3853 (1998).
63. S. Coulibaly, M. Taki, and N. Akhmediev, "Convection-induced stabilization of optical dissipative solitons," Opt. Lett. **36**(22), 4410-4412 (2011).
64. K. Staliunas, "Stabilization of spatial solitons by gain diffusion," Phys. Rev. A **61**(5), 053813 (2000).
65. P. V. Paulau, D. Gomila, P. Colet, B. A. Malomed, and W. J. Firth, "From one- to two-dimensional solitons in the Ginzburg-Landau model of lasers with frequency-selective feedback," Phys. Rev. E **84**(3), 036213 (2011).
66. N. V. Alexeeva, I. V. Barashenkov, K. Rayanov, and S. Flach, "Actively coupled optical waveguides," Phys. Rev. A **89**(1), 013848 (2014).
67. B. Dana, A. Bahabad. and B. A. Malomed, "CP symmetry in optical systems," Phys. Rev. A **91**(4), 043808 (2015).
68. J. Atai and B. A. Malomed, "Bound states of solitary pulses in linearly coupled Ginzburg-Landau equations," Phys. Lett. A **244**(6), 551–556 (1998).
69. G. Iooss and D. D. Joseph, *Elementary Stability and Bifurcation Theory* (Springer, 1980).
70. H. Uecker, D. Wetzel, and J. D. M. Rademacher, "pde2path - A Matlab package for continuation and bifurcation in 2D elliptic systems," arXiv:1208.3112 [math.AP], (2012).